\documentclass[hidelinks]{article}

\usepackage{arxiv}

\usepackage{graphicx}
\usepackage{xcolor}
\usepackage{cite}
\usepackage{subcaption}
\usepackage{booktabs}
\usepackage{siunitx}
\usepackage{hyperref}
\usepackage{listings}
\usepackage{wrapfig}
\usepackage{comment}

\newcommand{\orcid}[1]{\href{https://orcid.org/#1}{\includegraphics[height=10pt]{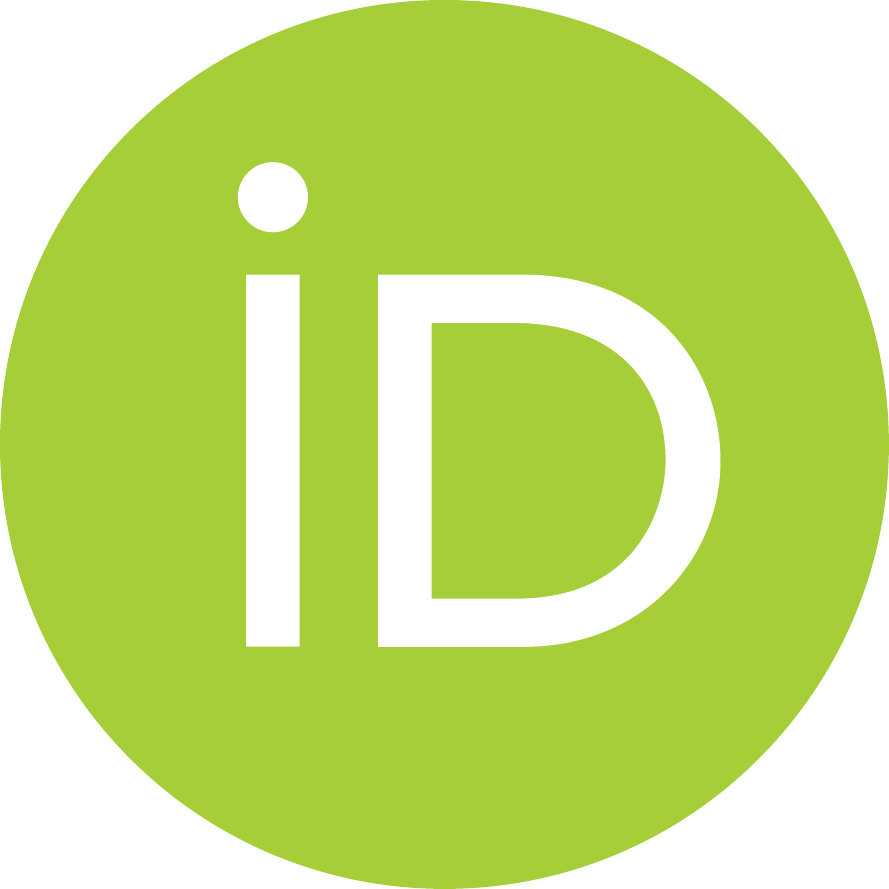}}}

\begin{document}
\title{Quantifying Overheads in Charm\texttt{++} and HPX using Task\ Bench}
%
%
\author{Nanmiao Wu\orcid{0000-0002-1825-7985}
\\
Center of Computation \& Technology, Lousiana State University\\
\{wnanmi1,igonid1,kmoham6,patrickdiehl,hkaiser\}@lsu.edu\\
Ioannis Gonidelis\\
Center of Computation \& Technology, Lousiana State University\\
Simeng Liu \\
Department of Computer Science, University of Illinois at Urbana-Champaign
\\
\{simengl2,zanef2,nikunj,kale\}@illinois.edu\\
Zane Fink\orcid{0000-0002-1882-6578} \\
Department of Computer Science, University of Illinois at Urbana-Champaign \\
Nikunj Gupta  \\ 
Department of Computer Science, University of Illinois at Urbana-Champaign\\
Karame Mohammadiporshokooh \\
Center of Computation \& Technology, Lousiana State University\\
Patrick Diehl\orcid{0000-0003-3922-8419} \\
Center of Computation \& Technology, Lousiana State University\\
Hartmut Kaiser\orcid{0000-0002-8712-2806} \\
Center of Computation \& Technology, Lousiana State University\\
Laxmikant V. Kale\orcid{0000-0001-9673-8445}\\
Department of Computer Science, University of Illinois at Urbana-Champaign
}

%
%
%
\maketitle              

\begin{abstract}
Asynchronous Many-Task (AMT) runtime systems take advantage of multi-core architectures with light-weight threads, asynchronous executions, and smart scheduling. In this paper, we present the comparison of the AMT systems Charm\texttt{++} and HPX with the main stream MPI, OpenMP, and MPI+OpenMP libraries using the Task\ Bench benchmarks. Charm\texttt{++} is a parallel programming language based on C\texttt{++}, supporting stackless tasks as well as light-weight threads asynchronously along with an adaptive runtime system. HPX is a C\texttt{++} library for concurrency and parallelism, exposing C\texttt{++} standards conforming API. First, we analyze the commonalities, differences, and advantageous scenarios of Charm\texttt{++} and HPX in detail. Further, to investigate the potential overheads introduced by the tasking systems of Charm\texttt{++} and HPX, we utilize an existing parameterized benchmark, Task\ Bench, wherein $15$ different programming systems were implemented, e.g., MPI, OpenMP, MPI + OpenMP, and extend Task\ Bench by adding HPX implementations. We quantify the overheads of Charm\texttt{++}, HPX, and the main stream libraries in different scenarios where a single task and multi-task are assigned to each core, respectively. 
We also investigate each system's scalability and the ability to hide the communication latency.

\keywords{Asynchronous Many-Task (AMT) \and Charm\texttt{++} \and HPX \and Task\ Bench.}
\end{abstract}
%
%
%
\section{Introduction}
Asynchronous Many-Task (AMT) systems emerge as an effective solution to the demands of adaptive applications. However, by utilizing the fine-grained parallelism, AMTs tend to generate runtime overheads which inhibits performance and counteracts their benefits. We are mainly interested in systems that expose distributed execution, which is the prevalent technique for massive computational experiments. Many options exist in the realm of parallel runtime systems, \emph{e.g.}\, Uintah~\cite{uintah}, Chapel~\cite{chapel}, Legion~\cite{legion}, and PaRSEC~\cite{parsec}. For a more detailed survey of various AMTs, we refer to~\cite{thoman2018taxonomy}. \textbf{This research focuses on Charm\texttt{++}~\cite{charm++} and HPX~\cite{kaiser2020hpx} since both systems provide a similar underlying programming model.}

Charm\texttt{++} delivers a highly abstracted environment for productivity bound to a flexible and performant execution paradigm. On the other hand, HPX provides a C\texttt{++} standards conforming API and extends the standard parallel facilities by providing asynchronous and distributed components. Our goal is to objectively quantify the overheads of those two systems by evaluating measurements from intrinsic benchmark implementations characteristics. Further comparisons are presented against MPI and OpenMP for distributed and within-node parallel execution, respectively.
For that, we utilize Task\ Bench, a unified benchmarking solution that evaluates these systems under a common ground. 

Task\ Bench was proposed by Slaughter et al.~\cite{10.5555/3433701.3433783} as a standardized solution that unifies the benchmarking process of various existing concurrency frameworks. It provides a backend benchmarking kernel that is exposed through a parameterized interface. Once Task\ Bench is implemented in a given programming system, it enables a straightforward comparative analysis with every other system in the Task\ Bench pool. Task\ Bench has already been implemented for Chapel~\cite{chapel}, Dask~\cite{dask}, MPI, OmpSs~\cite{ompss}, OpenMP, PaRSEC,
Realm~\cite{realm}, Regent~\cite{regent}, Spark~\cite{spark}, StarPU~\cite{starpu}, Swift/T~\cite{swiftt}, TensorFlow~\cite{tensorflow} and X10~\cite{x10}. This makes it suitable for our goal of a fair comparison of two different systems under a common ground. The centralized results enable a direct comparison of the performance of each system in a wide spectrum of tasking paradigms that model various execution schemes corresponding to real world experiments like the stencil pattern, the FFT pattern etc. We extend this work~\cite{10.5555/3433701.3433783} by adding HPX implementations, and reproducing the same results for Charm\texttt{++} and HPX with using MPI, OpenMP and MPI+OpenMP as a common denominator for the comparisons. Our results reflect those of the original authors and are accompanied by elaborate remarks. 

The three major contributions of our work are:
\begin{enumerate}
    \item This is the first work comparing Charm\texttt{++} and HPX using the same benchmark. Different HPX implementations with respect to the Task\ Bench library are implemented, namely HPX\ local and HPX\ distributed, in order to fairly compare HPX with Charm\texttt{++} against the mainstream MPI, OpenMP, and MPI+OpenMP. The optimizations of HPX implementations to minimize the overheads are further introduced.
    \item The commonalities, differences, and advantageous scenarios of Charm\texttt{++} and HPX, are analyzed in detail. The performance results further validate the analysis. 
    \item The overheads of Charm\texttt{++} and HPX, along with several other programming systems, are quantified in terms of shared-memory parallelism and distributed-memory parallelism, in various scenarios wherein a single task and multi-task are assigned to each core, respectively. 
\end{enumerate}

The paper is structured as follows: Section~\ref{sec:related:work} introduces the existing state-of-the-art performance evaluations. In Section~\ref{sec:amt} Charm\texttt{++} and HPX are briefly introduced and the similarities and differences are discussed. In Section~\ref{sec:taskbench} we briefly introduce the ingredients of Task\ Bench used in this work. Section~\ref{sec:improvements} summarizes the improvements to further reduce the overhead. Section~\ref{section:experiments} shows simulations in various scenarios without/with overdecomposition. Finally, Section~\ref{sec:conclusion} concludes the work.

\section{Related work}
\label{sec:related:work}
Many existing works evaluate the performance of task-based parallel programming models. In what follows, we consider studies focusing on mini-apps: simple applications designed to represent the performance characteristics of full-fledged applications.

Karlin et al.~\cite{Karlin2013:Exploring} evaluate the performance and productivity characteristics of several traditional and task-based parallel programming models using the LULESH~\cite{LULESH:versions} mini-application for shock hydrodynamics.
In~\cite{Rabbi2020:Evaluation}, the authors implement a block eigensolver in OpenMP~\cite{dagum1998openmp} and OpenACC~\cite{Wienke2012:OpenACC} to assess the performance portability of these models. 
The Parallel Research Kernels~\cite{VanDerWijngaart2014:Parallel,VanDerWijngaart2015:Using,VanDerWijngaart2017:New,VanDerWijngaart2016:Comparing} are a suite of mini-applications and microbenchmarks designed to assess the performance of different parallel systems and programming models. The authors in~\cite{Raut2021:Porting} implement a stencil mini-application in Legion~\cite{Bauer2012:Legion} and MPI, observing similar weak-scaling performance between Legion and MPI. In~\cite{Fink2021:Performance}, the mini-application and communication microbenchmark performance of Python ports of established programming models Charm\texttt{++} and MPI is compared.
An extensive study at Sandia National Laboratory \cite{baker2015asc} compares 3 many-task programming models on qualitative and quantitative metrics. 

While studies based on mini-application performance provide insight into the performance and programmability of different programming models, the $\mathcal{O}(m\cdot n)$ complexity of implementing $m$ mini-applications in $n$ frameworks makes it onerous to comprehensively evaluate even a few programming models on a range of benchmarks. Section~\ref{sec:taskbench} describes another approach to facilitate such comparisons.

\section{Asynchronous many-task systems}
\label{sec:amt}

\subsection{Charm\texttt{++}}
Charm\texttt{++} is a parallel programming language based on C\texttt{++}. Unlike the bulk-synchronous and process-centric approach taken by MPI, Charm\texttt{++} implements a migratable-objects programming model. The basic unit of object in Charm\texttt{++} is called a \textit{chare} which is typically a class in C\texttt{++}. Functions in a \textit{chare} can group logically-related execution and communication tasks, supporting data-encapsulation and locality. Users can designate some methods as \textit{entry methods} for a \textit{chare} class which are the methods that can be invoked by other, potentially remote, \textit{chares} asynchronously. With the object-oriented approach, Charm\texttt{++} supports overdecomposition, where the user can define multiple collections (``arrays") of \textit{chares} corresponding to the domain of the problem. Charm\texttt{++} applications typically partition the domain into finer grains than the amount of available execution units (\emph{e.g.}\ cores). The location of individual \textit{chares} is controlled dynamically by Charm\texttt{++}'s adaptive runtime system (aRTS). On each core (or node, in some configurations), a user-space scheduler is used to asynchronously but non-preemptively execute the set of available method invocations. This data-driven execution allows Charm\texttt{++} applications to adaptively overlap communication and computation. By leveraging migratability of \textit{chares}, the aRTS supports dynamic load-balancing, as well as other capabilities such as fault-tolerance, shrinking or expanding the set of nodes assigned to a job in the middle of execution, power/energy/thermal optimizations etc. 

\subsection{HPX}
HPX is a C\texttt{++} Standard Library for parallelism and concurrency~\cite{kaiser2020hpx}. HPX is implemented as a lightweight user-level task manager running on top of kernel threads. It is widely known that thread creation and destruction managed by the operating system are expensive and reserve lots of memory. For that reason, HPX creates one thread per core and binds each of them to one of the cores. Therefore, the performance can be improved since  there is no kernel-level interruption when the tasks are running. HPX is the first implementation of an advanced parallel execution model~\cite{hpx1}, which essentially resolves critical issues that prevent effective usage of new HPC systems: Starvation, Latency, Overheads, and Waiting for Contention. The HPX asynchronous programming model exposes a C\texttt{++} standard API entirely conforming to interfaces as defined by C\texttt{++}11/C\texttt{++}14/C\texttt{++}17/C\texttt{++}20 and adds on top of the latest C\texttt{++} standard by providing distributed and heterogeneous computing scenarios, which makes HPX portable and uniformly usable for local and remote parallelism. HPX aligns with the ongoing C\texttt{++} standardization proposal with a goal of providing a uniform interface, in particular, related to parallelism and concurrency. HPX is widely used for applications that utilize both shared and distributed memory. PeriHPX~\cite{diehl2020asynchronous} is an example of using HPX for shared-memory parallelism, and Octo-Tiger~\cite{marcello2021octo} is one example of using HPX for a distributed memory application.

\subsection{Commonalities  and differences}
Both Charm\texttt{++} and HPX are highly performant and feature rich AMTs that leverage asynchrony, overdecomposition, and migratability. These features are either implicitly or explicitly exposed to the user. For instance, Charm\texttt{++} supports built-in migrations while HPX implements user assisted migrations. Furthermore, they bring different interpretations and consequently implementation details on certain key concepts. Charm\texttt{++} defines a ``Processing Element" (PE) that can be an OS thread or a process. Each \textit{chare} is assigned to a PE by keeping it anchored to PEs to enhance locality of the computation. Note that \textit{chares} can move to another PE according to load-balancing strategies to minimize communication or achieve more balanced load distribution. Multiple \textit{chares} are assigned to a PE and user-level scheduler schedules entry method executions non preemptively based on availability of data (messages). On the other hand, HPX keeps the notion of locality explicit and the user needs to assign parallel execution to occur on a certain locality or locally if no locality is provided. Moreover, they both support threading, including features like thread suspension and resumption. While any parallel execution on HPX is run on an HPX thread, Charm\texttt{++} threading is mainly utilized only in specially designated threaded entry methods that use blocking primitives (such as access to \textit{futures}) that can otherwise block the scheduler if not run on a thread. The default entry methods are not threaded, and can be considered as stack-less tasklets. Finally, Charm\texttt{++} implements continuations by utilizing callbacks, while HPX utilizes C\texttt{++} conforming \textit{futures} that can retrieve the underlying computation result.

While both AMTs are feature rich, there are a few key areas in which Charm\texttt{++} is advantageous. As HPX utilizes HPX threads for any parallel execution, it suffers from the overheads of the threading subsystem and further overheads of the networking interface.
Charm\texttt{++} schedules over each PE individually, \emph{i.e.}\ anchoring each \textit{chares} to a particular PE (and thereby to a core) except when load-balancing, enhances locality and allows lock-less interaction between entities assigned to the same PE. Furthermore, Charm\texttt{++} supports load-balancing, automatic checkpoint-restart, and multiple communication protocols. Similarly, HPX supports load-balancing by enabling work-stealing scheduling policy, and supports explicit checkpoint, restart techniques, and several communication layers, \emph{e.g.}\ TCP, MPI, and libfabrics, with others currently under work.

HPX provides some clear advantages over Charm\texttt{++} as well. Given HPX exposes an ISO C\texttt{++} conforming API, porting any standard C\texttt{++} application to HPX is a mere search and replace. Porting to Charm\texttt{++} requires careful restructuring of the program. Furthermore, HPX supports all the C\texttt{++}17 parallel algorithms along with various execution policies. An application developer can use these execution policies to achieve NUMA aware parallelism, explicit vectorization of loops, asynchronous algorithm execution, and much more. Charm\texttt{++} requires the user to explicitly implement some of these features in their code. Given that HPX allows tracking of all function parameters and associated data either as a constant \textit{lvalue} reference or as \textit{rvalue} references, the overheads associated are minimal. Charm\texttt{++}'s parameter marshalling and related copying overheads, resulting in higher overheads in the single node shared-memory setting.

Thus, Charm\texttt{++} and HPX have similarities and differences, with multiple performance-oriented trade-offs based on the machine and programming model.

\section{Task\ Bench}
\label{sec:taskbench}
Task\ Bench is a parameterized benchmark for evaluating runtime system performance. Notably, Task\ Bench benchmarks are defined by task graphs expressing communication and task dependency patterns common in real-world applications.
This task graph representation enables the evaluation of $n$ systems for $m$ benchmarks with $O(m+n)$ implementation effort, rather than the $O(m\cdot n)$ effort required by other benchmark suites. This dramatically reduces the programming effort required to evaluate new systems and benchmarks. 




While \textit{strong} and \textit{weak} scaling have been the prevalent solutions for performance measurement, they both have the potential to yield misleading results. Strong scaling cannot isolate system overheads from application cost while weak scaling could hide the system overheads if large problem size is being used~\cite{10.5555/3433701.3433783}. Task\ Bench uses METG (\textit{Minimum Effective Task Granularity}) as a metric, which essentially indicates the scaling capabilities of the system on-target. METG exposes how high the computing performance (FLOP/s) can be maintained as the amount of work per task gets smaller. The reasoning behind METG is that for large problem sizes, all systems are expected to behave (almost) optimally. Conversely, for small problem sizes, parallelism becomes challenging. 
In this work, we use the same choice of $50$\% as the Task\ Bench paper~\cite{10.5555/3433701.3433783} to compare the smallest average task granularity such that each system reaches at least $50$\% peak efficiency. We briefly introduce the ingredients of Task\ Bench used in this paper. For more details, we refer to~\cite{10.5555/3433701.3433783}. 


\section{Improvements}
\label{sec:improvements}

\subsection{Charm\texttt{++}}\label{sub:charm_improvements}

Charm\texttt{++} has organically grown over 20+ years, along with many applications and research projects, such as fault tolerance and energy management. As a result, the most general implementation tends to have accumulated overheads. Especially for running fine-grained benchmarks, it is useful to select build-time options carefully. 
The following briefly describes relevant options:
\begin{itemize}
    \item \textbf{Eight-Byte Message Priority:} Charm\texttt{++} supports arbitrary-length bit-vector message priorities, complicating the message receive path. A build option
    to use eight-byte message priorities simplifies it. 
    
    \item \textbf{Simplified Scheduling Path:} We further simplify the message delivery path in Charm\texttt{++} with these additional changes: no message priorities, no idle detection, and no condition-based or periodic callbacks.
    
    \item \textbf{Intranode IPC via Shared Memory:} By default, Charm\texttt{++} uses the NIC for inter-process communication within a node. We assess the performance impact of shared-memory communication within a node.
\end{itemize}

While we use the Charm\texttt{++} implementation of Task Bench presented in~\cite{Slaughter2020:Task}, with the default build here, we provide some data with different build options to evaluate their impacts on fine-grained performance in section \ref{sub:charm_finegrain}.

\subsection{HPX}
For HPX, two implementations are available, one is HPX\ local and another is HPX\ distributed. Their similarity is that a scheduling facility that is based on top of the current C\texttt{++} Standard execution proposal~\cite{cppexecution}, called executor, is deployed on both of them. 
Utilizing such an executor, HPX implementations benefit from retaining the spawning threads alive by allocating existing work to these threads. Further, such executor offers more ability and flexibility, \emph{e.g.}\, users can determine the priority of worker threads, the stack size of the work threads, and enable or disable work-stealing policy. Note that work-stealing policy is advantageous when we consider overdecomposition, wherein each worker thread has a set of work in queue and the worker thread that finishes its local work can steal the work from currently active worker threads. HPX\ local and HPX\ distributed are also different. HPX\ local relies on HPX\ local facilities and does on-node computation, while HPX\ distributed depends on equivalent distributed facilities and manages communication on top of parallelization. 
\section{Experiments}
\label{section:experiments}

All experiments were conducted on Buran nodes of the Rostam cluster. The hardware and software details are shown in Table~\ref{rostam_info}. In Section~\ref{section:experiments:overhead} the overheads of Charm\texttt{++}, HPX, and other systems are measured when considering the scenario where the runtime overhead is dominant, and one computational task is assigned to each core. In Section~\ref{section:experiments:oversub}, overdecomposition is adopted where more than one  computational tasks are assigned to each core. We investigate the fine-grained performance of Charm\texttt{++} in Section~\ref{sub:charm_finegrain}, where we use POSIX shared memory for intra-node communication, as described in Section~\ref{sub:charm_improvements}. Each run is $1000$ time steps long. Each data point has run $5$ times, and a confidence interval with $99\%$ confidence level is shown for the variance in the $5$ runs.

\begin{table}[btp]
  \centering
  \caption{\textbf{Left} column: Compilers and libraries used to compile all systems. \textbf{Right} column: Hardware details of the rostam nodes.}
  \label{tab:nodes}
  \begin{tabular}{llll|ll}
    \toprule
    \multicolumn{4}{c|}{Software} & \multicolumn{2}{c}{Hardware}\\\midrule
    gcc       &  11.2.0  &  hwloc       &  2.6.0   & CPU &  AMD EPYC 7352 24-Core      \\ 
    boost       &  1.78.0  & OpenMPI      &  4.1.2   & Memory & 16 GB DDR-4 memory  \\  
    gperftools & 2.9.1 & cmake & 3.22.0 & Interconnect & 200Gb/s EDR Infiniband  \\
    \bottomrule
  \end{tabular}
  \label{rostam_info}
\end{table}

\subsection{Performance of a single task on each core}
\label{section:experiments:overhead}
To characterize the performance limited by runtime overhead, the number of tasks is set to the number of total cores. 

\begin{figure}[tbp]
    \centering
     \begin{subfigure}[b]{.45\textwidth}
    \centering
    \includegraphics[width=\linewidth,trim= 0 0 20 0]{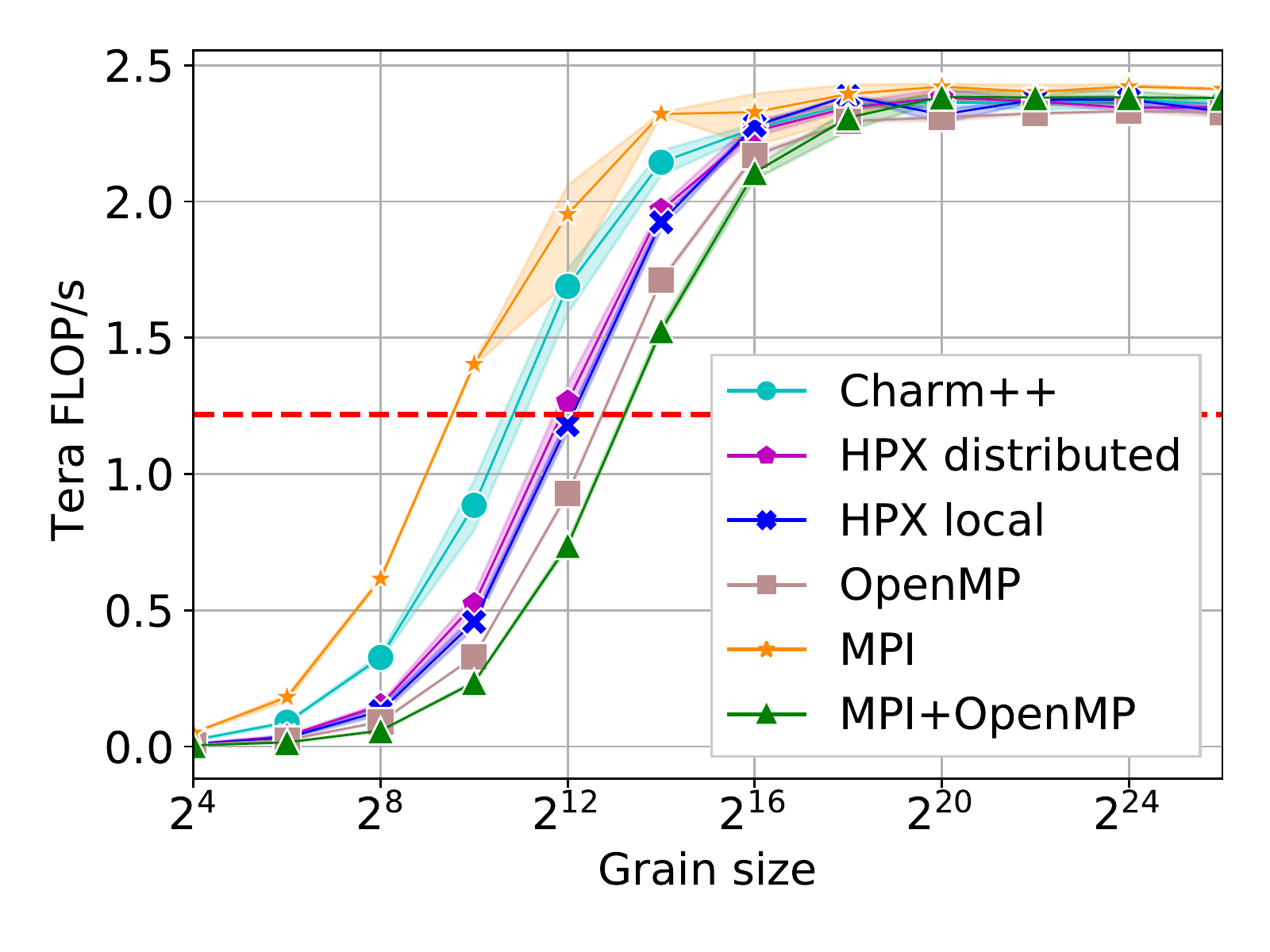}
    \caption{Tera FLOP/s vs grain size.}
    \label{fig:stencil_single_task_1node:a}
  \end{subfigure}
  \hfill
  \begin{subfigure}[b]{.45\textwidth}
   \centering
   \includegraphics[width=\linewidth,trim= 0 0 20 0]{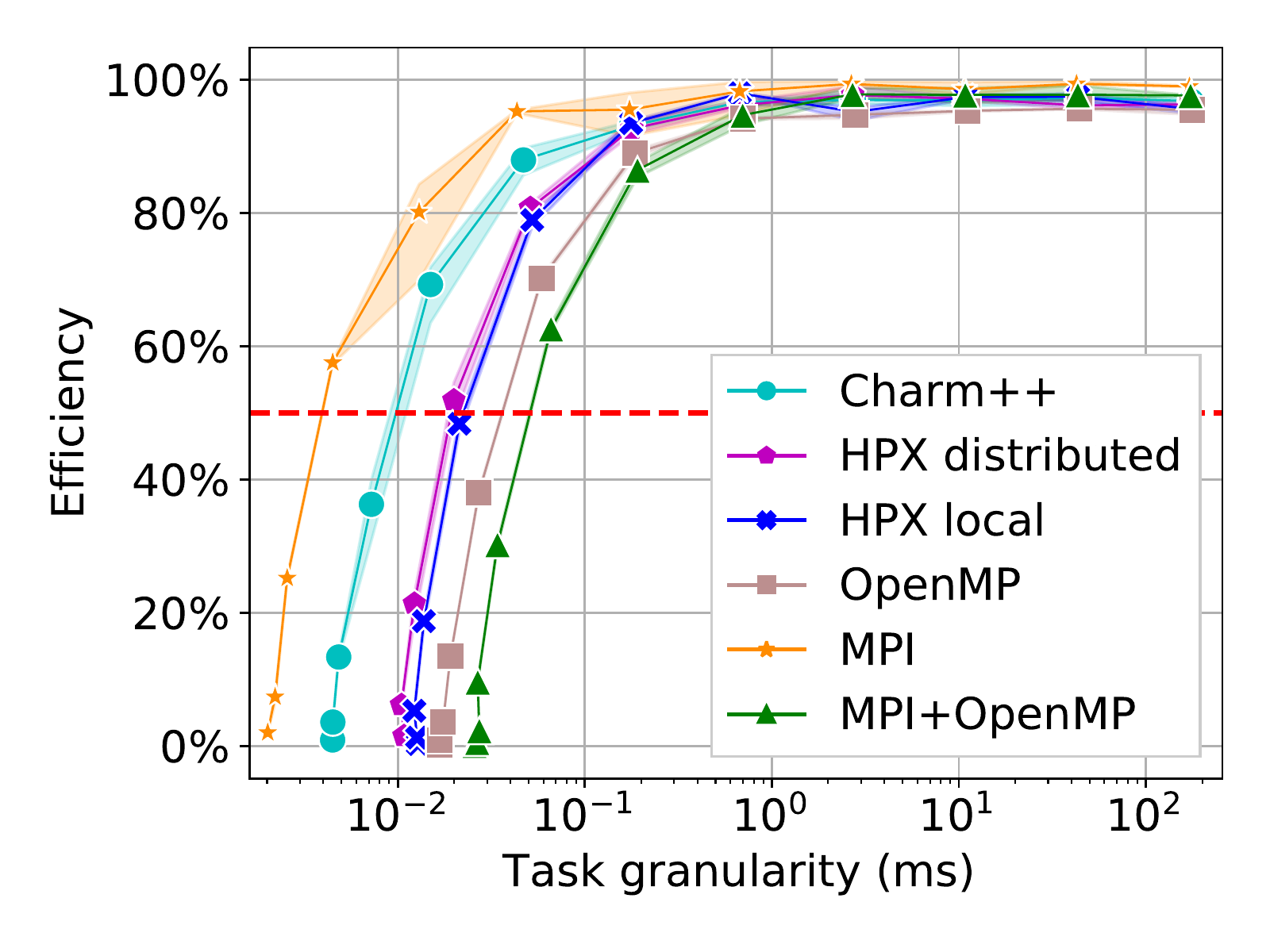}
  \caption{Efficiency vs task granularity}
  \label{fig:stencil_single_task_1node:b}
  \end{subfigure}
    \caption{Stencil pattern, $1$ node ($48$ cores), $48$ tasks.}
    \label{fig:stencil_single_task_1node}
\end{figure}

Figure~\ref{fig:stencil_single_task_1node:a} presents the TeraFLOP/s reached with a compute-bound kernel, varying the grain size. Note that the time for each vertex to execute such a kernel with a grain size of one is $2.5$ ns. Almost all systems achieve peak Tera FLOP/s, \emph{i.e.}\, $2.44\times 10^{12}$, when the grain size is large enough. Figure~\ref{fig:stencil_single_task_1node:b} shows the efficiency of each system responding
to the peak Tera FLOP/s vs. task granularity. Task granularity is measured by: wall time $\times$ number of cores $/$ number of tasks. Figure~\ref{fig:stencil_single_task_1node} shows METG of each system, which is the intersection of its efficiency curve and the $50$\% efficiency red dashed line in Figure~\ref{fig:stencil_single_task_1node}. To calculate METG, we first measure the peak Tera FLOP/s and get the efficiency percentage of each system responding to the peak Tera FLOP/s. For more details about METG, we refer to~\cite{10.5555/3433701.3433783}. For the shared-memory system, \emph{i.e.}\, OpenMP and HPX\ local, we observe that HPX\ local performs better than OpenMP.
For the distributed-memory system, 
we find that MPI has the smallest METG, $3.9$ $\mu$s. METGs of other systems for this scenario are listed in the first column of Table~\ref{tab:metg_stencil_1node}.

\begin{table}[tbp]
\caption{METG ($\mu$s) of each system for the stencil pattern without/with different overdecomposition, using 1 node. }
\label{tab:metg_stencil_1node}
  \centering
\begin{tabular}{l| c c | c c | c c} 
 \toprule
 System  & single task per core &  & 8 tasks per core &  & 16 tasks per core  \\ [0.5ex] 
 \midrule
 Charm\texttt{++} & \num[round-mode=places, round-precision=1]{9.7605} & &  \num[round-mode=places, round-precision=1]{37.8166} & & \num[round-mode=places, round-precision=1]{84.078} \\ 
 HPX\ distributed &  \num[round-mode=places, round-precision=1]{19.3313} & & \num[round-mode=places, round-precision=1]{39.2331} &  &\num[round-mode=places, round-precision=1]{54.0927} \\
 HPX\ local & \num[round-mode=places, round-precision=1]{22.3684} &  & \num[round-mode=places, round-precision=1]{54.4945} & & \num[round-mode=places, round-precision=1]{77.9015} \\
 MPI & \num[round-mode=places, round-precision=1]{3.92963} & & \num[round-mode=places, round-precision=1]{6.1046} & & \num[round-mode=places, round-precision=1]{7.64} \\
 OpenMP & \num[round-mode=places, round-precision=1]{36.178} & & \num[round-mode=places, round-precision=1]{36.9098} & & \num[round-mode=places, round-precision=1]{41.7562} \\
 MPI+OpenMP & \num[round-mode=places, round-precision=1]{50.8731} & & \num[round-mode=places, round-precision=1]{152.51} & & \num[round-mode=places, round-precision=1]{258.571} \\
 \bottomrule
\end{tabular}
\end{table}

\subsection{Performance of overdecomposition}
\label{section:experiments:oversub}
To quantify the performance of overlapping communication with computation, the total size of tasks is set to $N$ times the number of total cores, such that each core processes $N$ tasks. In this subsection, $N$ is set to $8$ and $16$, respectively. 

Table~\ref{tab:metg_stencil_1node} lists METGs of each system for the stencil pattern with/without overdecomposition,
using one node, respectively. For all systems, MPI achieves the smallest METG for these three scenarios. 

\begin{figure}[tbp]
    \centering
     \begin{subfigure}[b]{.45\textwidth}
   \centering
   \includegraphics[width=\linewidth,trim= 0 0 20 0]{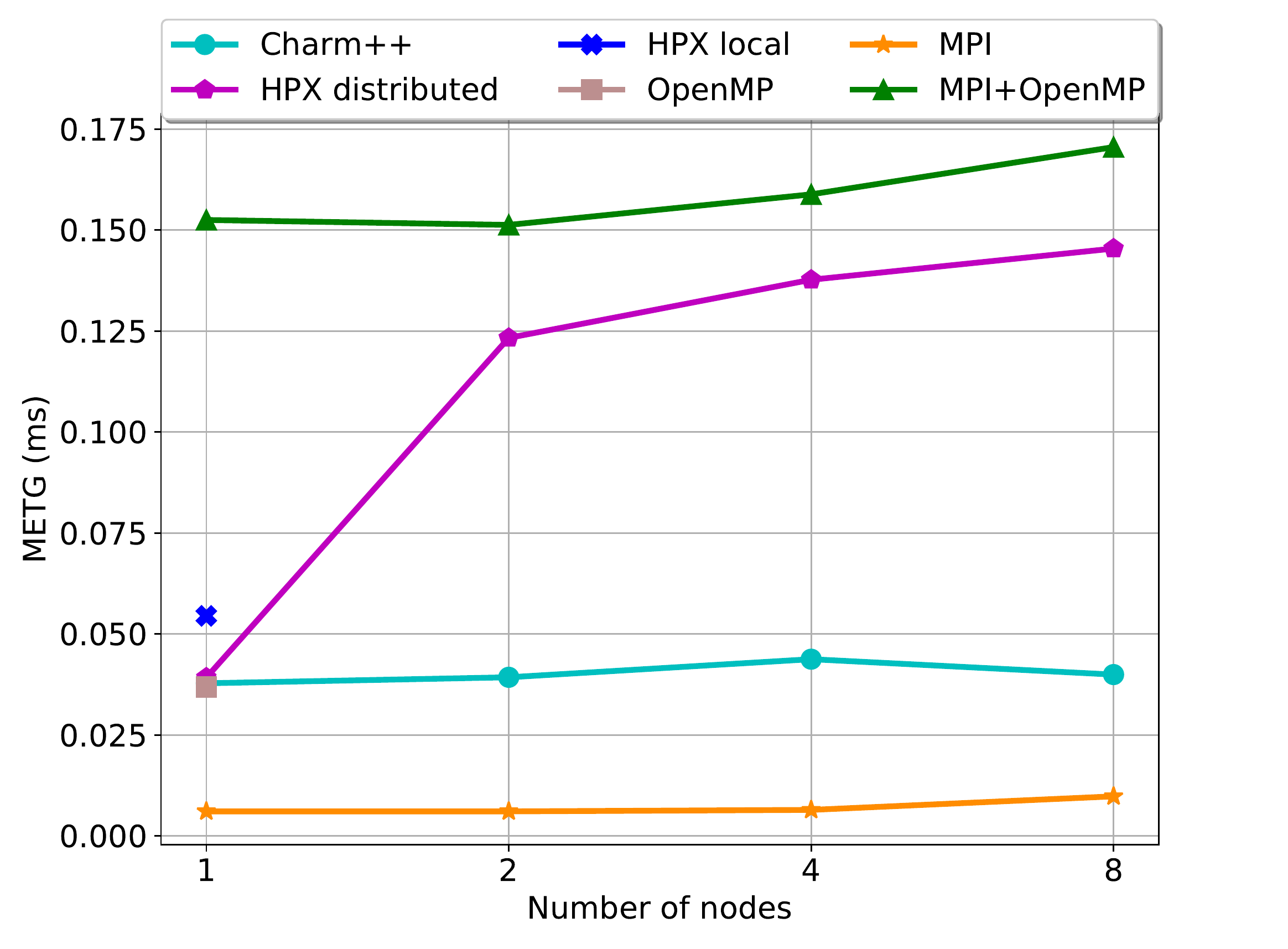}
  \caption{Stencil pattern, overdecomposition $8$ ($8$ tasks per core).}
  \label{fig:stencil_metg_odf8}
   \end{subfigure}
  \hfill
  \begin{subfigure}[b]{.45\textwidth}
    \centering
    \includegraphics[width=\linewidth,trim= 0 0 20 0]{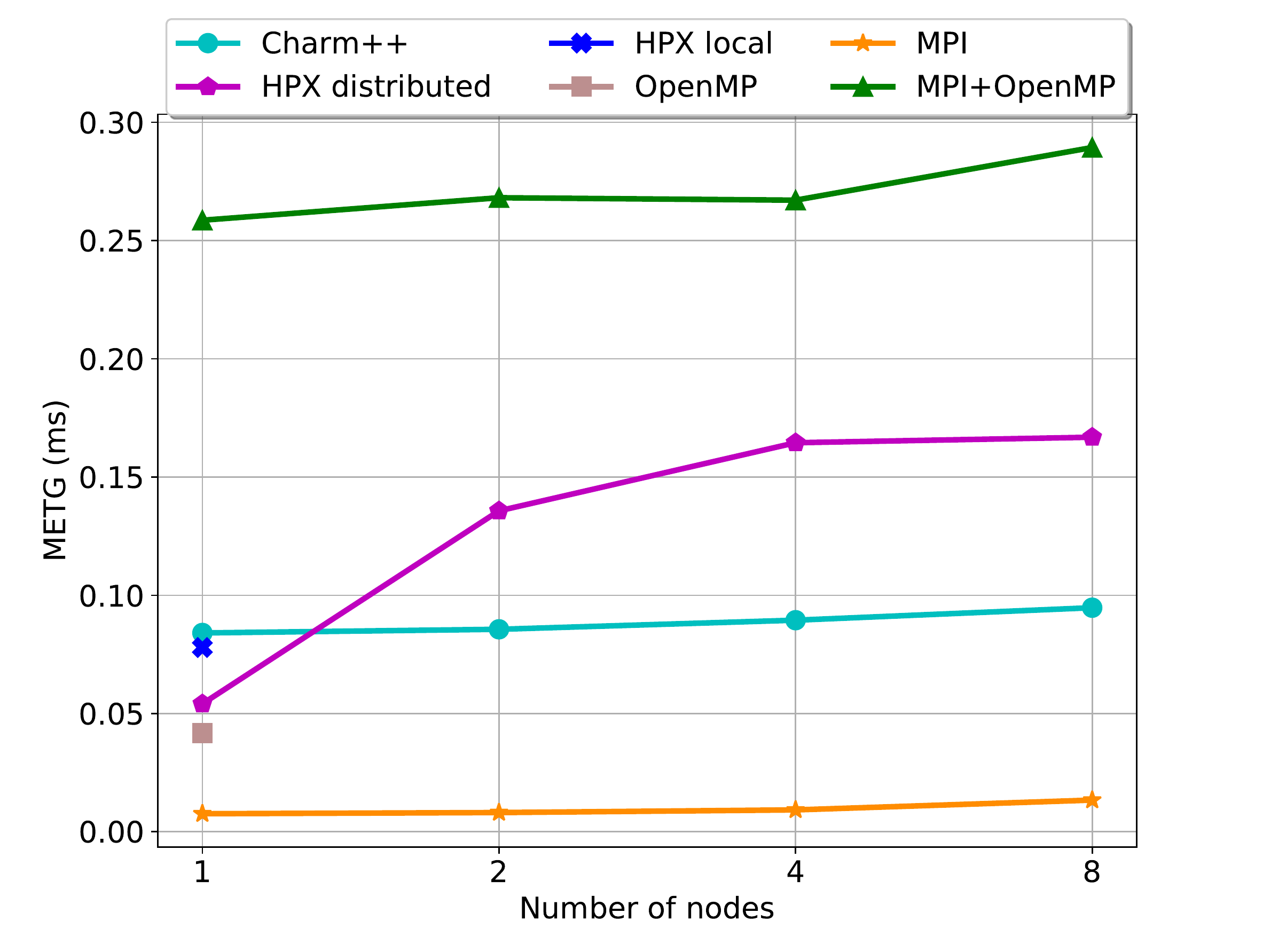}
    \caption{Stencil pattern, overdecomposition $16$ ($16$ tasks per core).}
    \label{fig:stencil_metg_odf16}
  \end{subfigure}
   \label{fig:stencil_metg_odf}
    \caption{METG of each system with varying number of nodes for different overdecomposition. METG is short for Minimum  Effective  Task  Granularity, is an efficiency-constrained metric for runtime-limited performance, introduced in Task\ Bench paper~\cite{10.5555/3433701.3433783}.} 
    \label{fig:metg}
\end{figure}

Figure~\ref{fig:metg} presents METGs of each system with varying number of nodes. Lower is better because a lower METG indicates a smaller task granularity required to achieve at least $50\%$ overall efficiency. Flat is ideal because a flat line implies that the communication topology does not affect METG by increasing the number of nodes. We observe that Charm\texttt{++} and MPI have lower and flat trends, while HPX\ distributed and MPI+OpenMP have higher and rising tendencies. For shared-memory parallelism, OpenMP has smaller METGs than HPX\ local for both scenarios. 

\vspace*{0.2in}
\subsection{Fine-grained Charm\texttt{++} performance}\label{sub:charm_finegrain}

\begin{figure}[tpb]
  \begin{center}
     \includegraphics[width=0.35\linewidth]{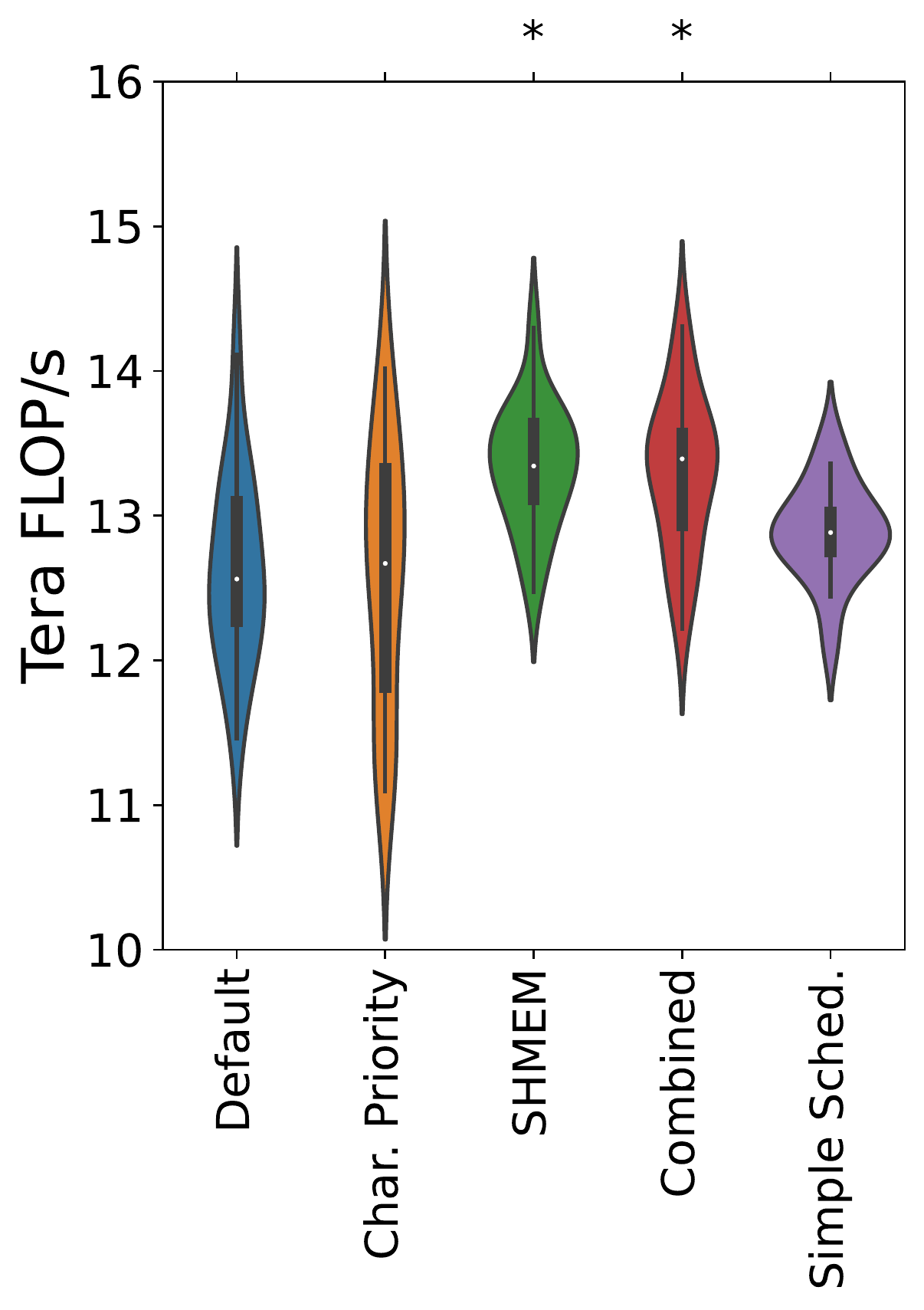}
      \end{center}
    \caption{Stencil pattern, $8$ nodes, $384$ cores, $384$ tasks. Performance of different Charm\texttt{++} build configurations for a grain size of 4096 iterations.} 
    \label{fig:charm_fine_grain}
\end{figure}

In Figure~\ref{fig:charm_fine_grain}, we evaluate the performance impact of the different build options meant for fine-grained applications described in Section~\ref{sub:charm_improvements}, which were not used in the above experiments. \texttt{Default} is the standard Charm\texttt{++} build used above. 
\texttt{Char. Priority} denotes a build using eight-byte message priorities;  \texttt{SHMEM} denotes the build that uses shared-memory for intra-node communication. \texttt{Combined} is a build using all optimizations, and \texttt{Simple Sched.} denotes Charm\texttt{++} built with the simplified scheduling path described in Section~\ref{sub:charm_improvements}.

We find that \texttt{SHMEM} and \texttt{Combined} yield an average throughput increase of $5.7\%$ and $5.3\%$, respectively
Using eight-byte message priorities and a simplified scheduling path did not yield a significant increase in throughput. Consequently, we find that scheduling overhead is not substantial even at this grain size, and that communication latency dominates. To further explore the performance impact of scheduling, additional investigation with different Task Bench dependency patterns is required.



\section{Conclusion and outlook}
\label{sec:conclusion}
This work is the first work comparing Charm\texttt{++} and HPX using the same benchmark. 
Using Task\ Bench enabled us to study the overheads introduced by the two AMTs compared to the more traditional approaches. The asynchronous scheduling using light-weight threads as in HPX or stackless tasks as in Charm\texttt{++} incurred some costs. We seen, for larger grain sizes, the overhead was negligible. However, for smaller grain sizes, the overhead was observed. To conclude, the overheads of fine-grained parallelism was not inherent to the programming models, and benchmark studies like this one was expected to lead to further optimizations to reduce or eliminate the gap with respect to MPI. 

This study has shown that there is potential for improvement for both AMTs for smaller grain sizes. Here, we need to investigate the differences with respect to MPI and do some profiling with the tools provided by both AMTs. For distributed HPX, we plan to try different libraries for communication, \emph{e.g.}\ libfabric and LCI. For Charm\texttt{++} the support for active messaging in the communication layer (such as UCX) will be tested. As a next step, a comparison with other AMTs would be interesting. 

\bibliographystyle{IEEEtran}
\bibliography{ref.bib}
\end{document}